\documentclass[letterpaper,twocolumn,10pt]{article}
\usepackage{usenix2020_SOUPS}

\usepackage{tikz}
\usepackage{amsmath}
\usepackage{multicol}

\begin{document}

\date{}

\title{\Large \bf Multi-armed bandit approach to password guessing}

\def\plainauthor{Hazel Murray and David Malone}

\author{
{\rm Hazel Murray}\\
Maynooth University\\Ireland\\hazel.murray@mu.ie
\and
{\rm David Malone}\\
Maynooth University\\Ireland\\david.malone@mu.ie
} 

\maketitle
\thecopyright

\begin{abstract}

The multi-armed bandit is a mathematical interpretation of the problem a gambler faces when confronted with a number of different machines (bandits). The gambler wants to explore different machines to discover which machine offers the best rewards, but simultaneously wants to exploit the most profitable machine. A password guesser is faced with a similar dilemma. They have lists of leaked password sets, dictionaries of words, and demographic information about the users, but they don't know which dictionary will reap the best rewards. In this paper we provide a framework for using the multi-armed bandit problem in the context of the password guesser and use some examples to show that it can be effective. 
\end{abstract}

\section{Introduction}
Passwords are a widely used form of authentication online. However, one major weakness of passwords is that human chosen passwords can be easily guessed by automated attacks. In fact, with the regular occurrence of leaks of password datasets, attackers are provided with an increasing amount of data to inform password guesses. It is important for security advocates and researchers to understand the capabilities of attackers given they have access to this data. This way, we can create countermeasures to protect the security of Internet users. 

Users from similar demographics will often choose similar passwords \cite{malone2012investigating,murray2020convergence}. In addition, users have been observed choosing passwords that reflect the nature of the website they are choosing the password for \cite{wei2018password}. In this work, we investigate whether an automated learning algorithm can identify these idiosyncrasies within a password set and whether it can leverage this knowledge in order to improve the success of password guessing rates.

A commonly used method for guessing passwords involves using dictionaries of password guesses and guessing these in an optimum order in order to compromise as many users as possible.
Guessing passwords in the optimal order is important for an attacker as they wish to compromise many users with a small number of guesses. In particular, an online attacker is often using automated attacks and is limited to a certain number of guesses before a lockout is triggered. It is a challenge for an attacker to discover which guesses will result in the highest success rates.

In this paper we suggest and develop an \textit{explore and exploit} protocol based on the classic multi-armed bandit problem.  This protocol can be used to guess a password set effectively using guesses from a selection of dictionaries. 

In Section~\ref{sec:related work}, we describe related work. In Section~\ref{sec:mab_prob1}, we describe the multi-armed bandit problem in the context of password guessing and the iterative methods used to solve it. In Section~\ref{sec:mab_model2}, we describe the implementation of the bandit model and in Section~\ref{sec:mab_results} we provide the results of this implementation for dictionaries of leaked password sets. Finally, in Section~\ref{sec:conclusion} we summarise our results and plan future work.

\section{Related work}\label{sec:related work}
For a long time researchers have been interested in modelling and improving password guessing. The first strategic methods involved dictionary attacks. There were proposed by Morris and Thompson in 1979~\cite{morris1979password} and are still widely used today~\cite{jtr,hashcat}. In 2005, Narayanan et al. employed Markov models to enable faster dictionary attacks~\cite{narayanan2005fast}. In 2009, Weir et al. used probabilistic context-free grammar (PCFG) which were trained using password breaches and used to assign probabilities to passwords for guessing~\cite{weir2009password}. In 2013, D\"{u}rmuth et al. proposed an updated password guessing model based on Markov models, called OMEN~\cite{castelluccia2013privacy}. As part of their initial paper they demonstrated an OMEN specific method for merging personal information with a dictionary of guesses. They acknowledged the difficulty of merging guesses from two different sources. In 2016, Wang et al. developed a targetted password guessing model which seeds guesses using users' personally identifiable information. In 2017, Houshmand and Aggarwal created a method for merging multiple grammars for dictionary-based PCFG models~\cite{houshmand2017using}. Pal et al. in 2019, developed a password manipulation tool called PASS2PATH. Leveraging the knowledge that users alter and reuse their passwords, this guessing model can transform a base user password into effective targeted password guesses. 

Determining the order to make password guesses is a hard problem. Particularly when drawing from multiple sources which often do not provide obvious probability scores. We imagine an attacker has multiple sources of information for informing password guesses: previous password leaks, personal and demographic information about the users, and simple dictionary words, among other things. In this paper, we describe a method for optimising the choice of guesses by combining the information from all the available dictionaries. This is a new piece of research that, to the best of our knowledge, has not been tackled analytically.




\section{Multi armed bandit problem}\label{sec:mab_prob1}
The multi-armed bandit problem describes the trade-off a gambler faces when faced with a number of different gambling machines. Each machine provides a random reward from a probability distribution specific to that machine. The crucial problem the gambler faces is how much time to spend \textit{exploring} different machines and how much time to spend \textit{exploiting} the machine that seems to offer the best rewards.
The objective of the gambler is to maximize the sum of rewards earned through a sequence of lever pulls.

In our scenario, we regard each dictionary as a machine which will give a certain distribution of successes. We want to explore the returns from each dictionary and also exploit the most promising dictionary, in order to make effective guesses. With each guess we learn more about the distribution of the password set we are trying to guess. Leveraging this knowledge, we can guess using the dictionary that best matches the password set distribution, thus maximising rewards. 



In the following sections we describe the set up of the multi-armed bandit problem in the context of password guessing. 

\subsection{Problem set up}\label{sec:formal prob}
Suppose we have $n$ dictionaries. Each dictionary $i = 1 \ldots n$, has a probability distribution $p_i$, and $\sigma_i(k)$ denotes the position of password $k$ in dictionary $i$. So, the probability assigned to password $k$ in dictionary $i$ is $p_{i,\sigma_i(k)}$.

Suppose we make $m$ guesses where the words guessed are $k_j$ for $j = 1 \ldots m$. Each of these words is guessed against the $N$ users in the password set and we find $N_j$, the number of users' passwords compromised with guess number $j$.

To model the password set that we are trying to guess, we suppose it has been generated by choosing passwords from our $n$ dictionaries. 
Let $q_i$ be the proportion of passwords from dictionary $i$ that generated the password set. Our aim will be to estimate $q_1, \cdots, q_n$ noting that
\begin{equation} \sum_i^n q_i = 1 \qquad \mbox{and} \qquad q_i \geq 0. \label{eq:probconstraint}
\end{equation}

If the password set was really composed from the dictionaries with proportions $q_i$, the probability of seeing password $k$ in the password set would be
\begin{equation}
Q_k := \sum_{i=1}^{n} q_i p_{i,\sigma_i(k)}.
\label{eq:passwordproportion}
\end{equation}
Given the $N_j$, we will use this probability to build a maximum likelihood estimator.

\subsection{Likelihood estimator}\label{sec:likelihood}
Maximum likelihood estimation is a method for estimating the parameters of a probability distribution using observed data. It does so by selecting the parameters so that the observed data is most probable. 

We construct the following likelihood for our model with $m$ guesses: 
\begin{align*}
    \mathcal{L} =  & \binom{N}{N_1 \,  \cdots \, N_m \, (N - N_1 \cdots -N_m)} Q_{k_1}^{N^1} Q_{k_2}^{N^2}  \cdots Q_{k_m}^{N^m}\\
    &\hspace{1em} \cdot \left(1- Q_{k_1} \cdots - Q_{k_m} \right)^{N -N_1 \cdots - N_m}
\end{align*}


Our goal is the maximise this likelihood function by choosing good estimates for $q_1, \ldots q_n$ based on our observed rewards from each previous guess.  Note, with each guess we learn more about $q_i$ for all the dictionaries. In fact, one of the features of this model compared to a traditional multi-armed bandit model is that when we make a guess we learn something about all the dictionaries.


\subsection{Maximising the likelihood function}\label{sec:max}
Except in limited cases, the likelihood equation cannot be solved explicitly. We choose to use gradient descent to find the $q_i$ that maximise $\mathcal{L}$ after $m$ guesses subject to the constraints in (\ref{eq:probconstraint}). We were able to prove that the log-likelihood function for our system is concave, and therefore the likelihood function has a unique maximum value.

We iteratively changed the estimated $q_i$ values, $\hat q_i$. To ensure we met the constraints in (\ref{eq:probconstraint}), we project the gradient vector onto the probability simplex and then constrain our steps so that we always stay within that space. 



Using this gradient descent method, we should be able to estimate the parameters $q_1, \cdots, q_n$ for each dictionary based on the information we have collected from previous guesses.


\subsection{Gradient descent results}
The goal of the gradient descent is to converge towards the maximum of the likelihood function and thus find the proportions $q_i$ that provide the best explanation of the distribution of the password set seen after $m$ guesses.

To determine whether this is working we take four different leaked password datasets as dictionaries. The datasets we used are leaked passwords from users of hotmail.com, flirtlife.de, computerbits.ie and 000webhost.com. They contain $7300$, $98912$, $1795$ and $15,252,206$ users' passwords respectively\footnote{The datasets  were compromised by various methods so the lists may only contain a random, and possibly biased, sample of users~\cite{murray2020convergence}. As far as we can tell only the 000webhost dataset imposed composition policies on users~\cite{golla2018accuracy}.}.

We took a random sample of 1000 users' passwords from the 98912 users in the flirtlife dataset. In previous work, we showed that when guessing a sample of leaked passwords from a website, the most effective guesses will come from the passwords of other users of that same site~\cite{murray2020convergence}. Therefore, if the multi-armed bandit is effective we expect it to show that the sample most closely compares to the flirtlife dictionary. 

In Figure~\ref{fig:flife,descent}, we show the $q$ value estimates during the convergence of the gradient descent for the likelihood function seeded with just one guess. This single guess was the password \textit{123456}, widely considered the most commonly used password. 
We began by setting the $\hat q$-values to $1/n = 0.25$, and then used 100 steps of gradient descent to estimate the proportions. As expected, the estimation suggests that a high proportion of the passwords in the sample were drawn from the flirtlife.de dictionary.

The password \textit{123456} occurred in all four password sets but using the distribution of those datasets the likelihood function was able to determine that the proportion in the sample best matched the proportion in flirtlife. If we were guessing the full flirtlife dataset with several guesses, rather than just a sample from it with one guess, then this proportion would be closer to 100\%. 


\begin{figure}
    \centering
    \includegraphics[width=1\columnwidth]{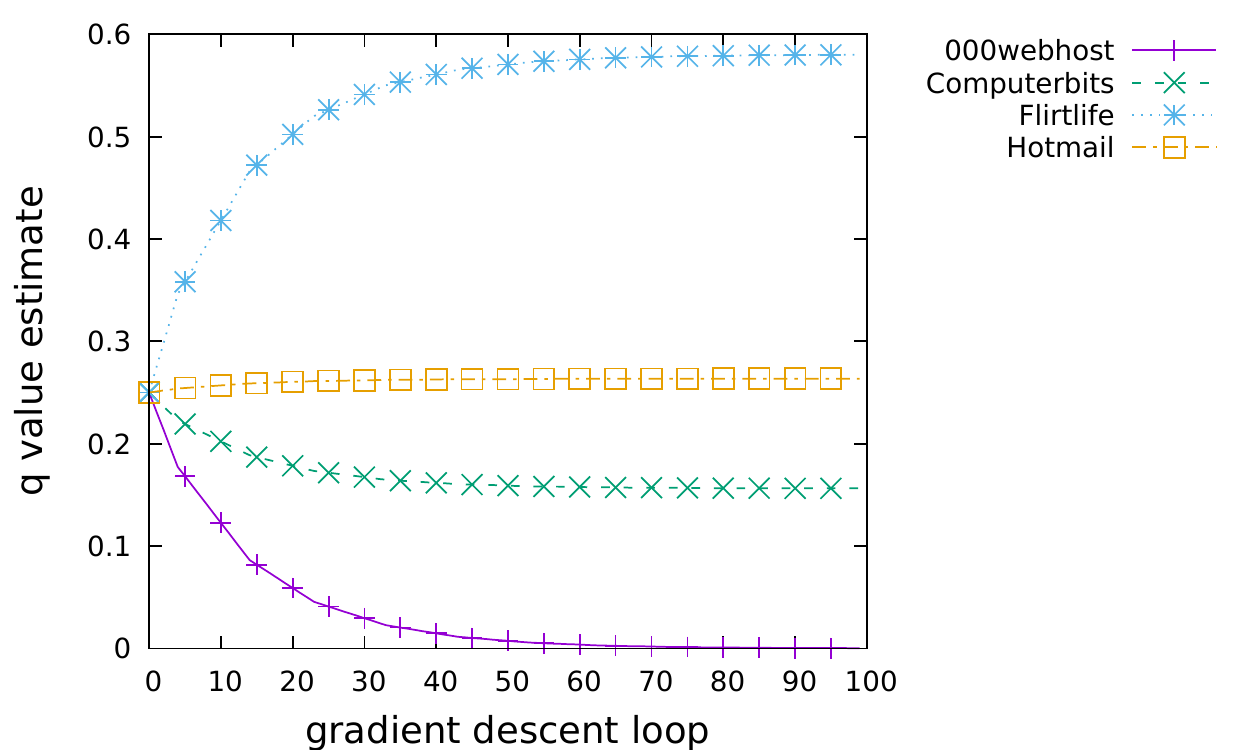}
    \caption{Estimating the distribution of passwords using information from 1 guess.}
    \label{fig:flife,descent}
\end{figure}

The optimization assigns a non-zero proportion to two of the other dictionaries. This implies these dictionaries hold some guessing value. As we collect more information from additional guesses the proportion assigned to those dictionaries will converge to zero.

In the above example, all 1000 passwords came from a random sample of flirtlife. We will investigate in later examples (e.g. Figure~\ref{fig:q-pwdset1}, Figure ~\ref{fig:q-pwdset2} and Figure ~\ref{fig:q-pwdset4}) whether the maximum likelihood estimation can determine the breakdown of where passwords come from when composed of different dictionaries.

\section{Multi-armed bandit model}\label{sec:mab_model2}



Let us now suggest some variations of the gradient descent and the maximum likelihood estimates when forming a password guessing multi-armed bandit. We are interested in which of these variations produces the best results.

\subsection{Gradient descent initialization variables}
We expect the gradient descent to improve with each guess made since every guess allows it to gain more information. There are a number of ways of initialising the gradient descent after each guess provides new information. The following are three different methods for choosing the initialisation value:
\begin{description}
    \item[Random] Randomly pick starting values for $\hat q_i$, subject to (\ref{eq:probconstraint})
    \item[Average] Choose the average starting value, i.e. assume the passwords are distributed evenly between the $n$ dictionaries, so $ \hat q_i = 1/n$
    \item[Best] Use our previous best estimate for the $\hat q$-values, based on the gradient descent results for the previous guess.
\end{description}

\subsection{Informing our guess choices}
Once we have generated our estimate of the $\hat q$-values, we want to use them to inform our next guess. We suggest three options for how to choose our next guess:
\begin{description}
    \item[Random] Randomly choose a dictionary and guess the next most popular password in that dictionary.
    \item[Best dictionary] Guess the next most popular password from the dictionary with the highest corresponding $\hat q$-value.
    \item[By Q] Use all the information from all the $\hat q$-values for the dictionaries and all the frequencies of the passwords in the dictionaries to inform our next guess.
\end{description}

These options have different advantages. In the first option, we randomly choose a dictionary to guess from, but we are still taking the most probable guess from the dictionary we choose. This option emphasises the continued exploration of all the dictionaries. In the second option, we are choosing the dictionary we believe accounts for the largest proportion of the password set.

The last option is specifically basing password guess choices on equation (\ref{eq:passwordproportion}). It uses our predicted $\hat q$-values to estimate the probability of seeing each word $k$. If, for example, we have a word $k$ which has frequency $f_1(k)$ in dictionary 1 but also occurs in dictionary 2 and 3 with frequencies $f_2(k)$ and $f_3(k)$ respectively. Using the above computation, where $p_{i,\sigma_i(k)}= f_i(k)/\mbox{\textit{size of dictionary }} i $, we can compute the total probability of this word occurring in the password set. This method should determine which word $k$ has the highest probability of being in the password set and use this word as our next guess. 



\section{Multi-armed bandit Results}\label{sec:mab_results}
We will now look at some examples of the performance of our multi-armed bandit model. It will guess one word at a time and then compute the estimated weighting of each dictionary. We will vary the composition of the password set and consider some of the variations described in Section~\ref{sec:mab_model2}. 

\subsection{Password set 1: 60\% flirtlife, 30\% hotmail, 10\% computerbits}
We begin by guessing a password set made up of 10,000 users' passwords; 60\% were selected randomly from flirtlife, 30\% from hotmail and 10\% from the computerbits users.

\begin{figure}
    \centering
    \includegraphics[width=1\columnwidth]{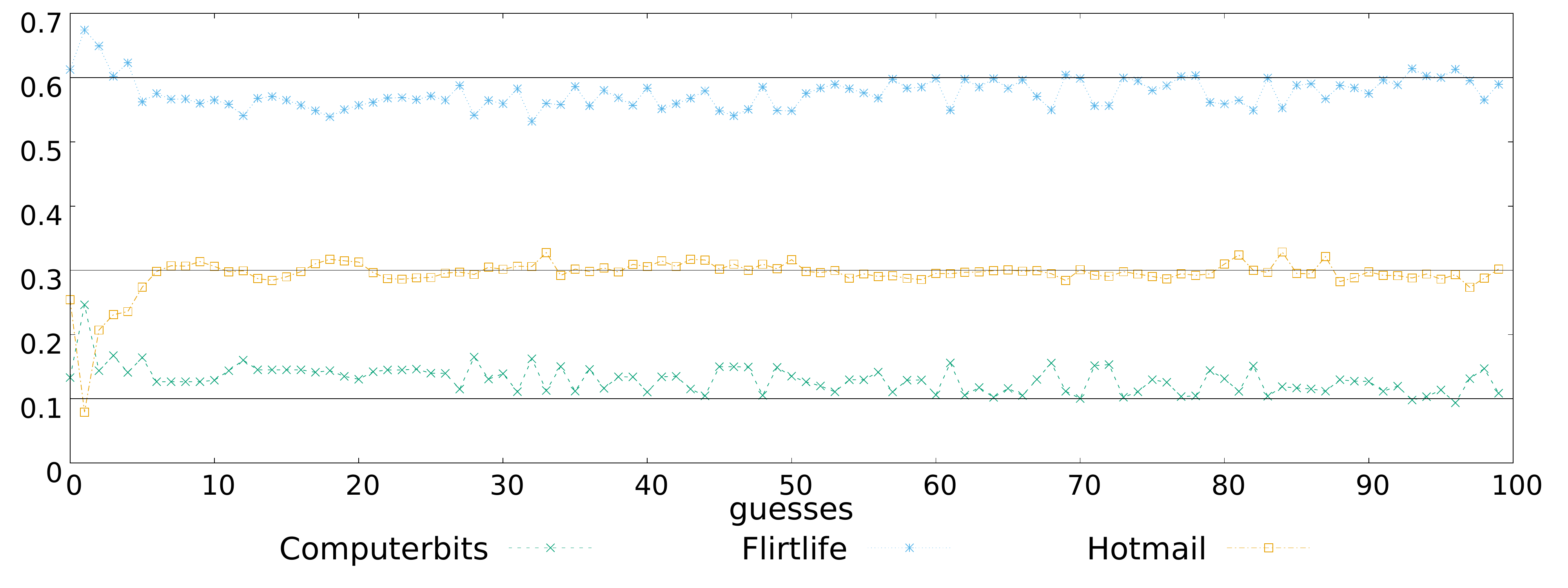}
            \vspace{-0.3cm}
    \caption{Password set 1 $q$-value estimates. Initialization: average $\hat q$-values, Guessing: by $\mathcal{Q}$.}
    \label{fig:q-pwdset1}
    \vspace{-0.3cm}
\end{figure}

In Figure~\ref{fig:q-pwdset1}, we plot the estimated $q$-values after the gradient descent was completed for each guess. For this graph, the gradient descent was initialised using average $\hat q$-values, $\hat q_i = 1/3$, and the $\mathcal{Q}$ method was used for guessing. The actual proportions are shown as solid horizontal lines. Even after a small number of guesses we have good predictions for how the password set is distributed between the three dictionaries. 


\begin{figure}
    \centering
    \includegraphics[width=1\columnwidth]{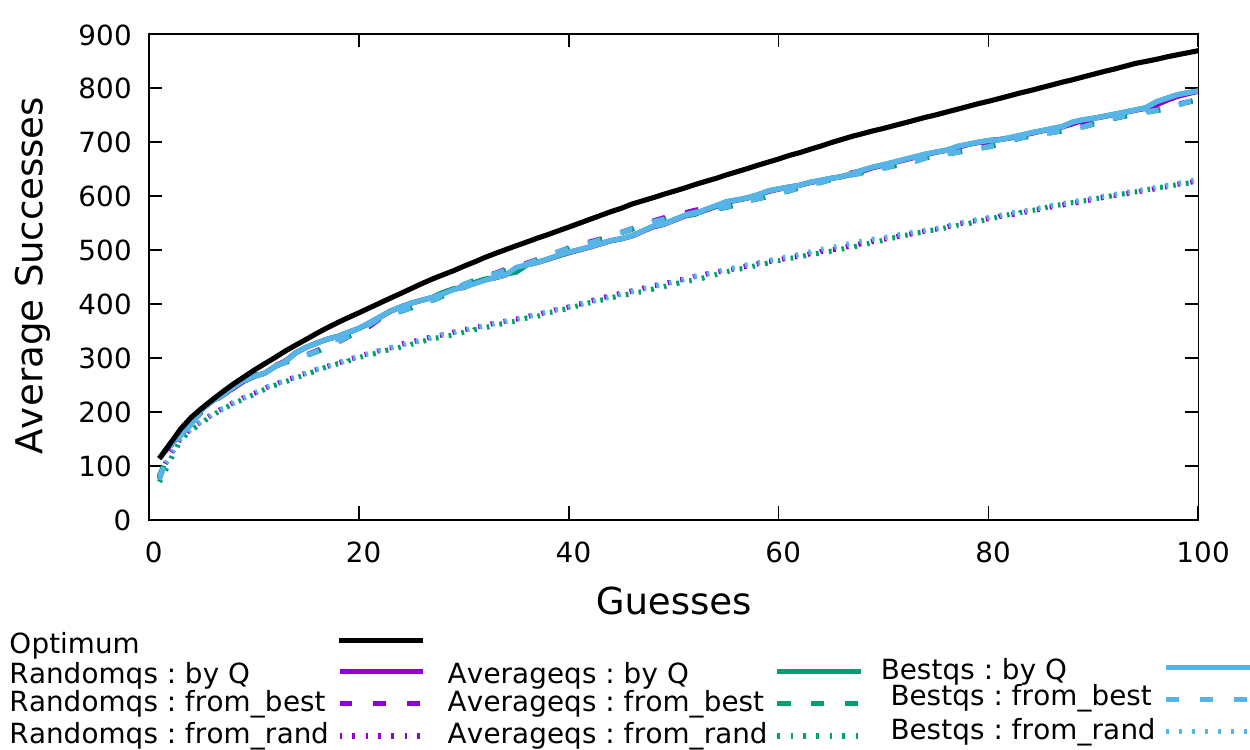}
    \caption{Guessing returns for password set 1.}
    \label{fig:success-pwdset1}

\end{figure}

In Figure~\ref{fig:success-pwdset1}, we show  the number of users successfully compromised as the number of guesses increases. The successes are the average over fifty runs to reduce the variance in the random guessing method. Results are shown for each combination of initialisation and guessing method. As one might expect, picking guesses from a dictionary at random resulted in the lowest success rates. Both the $\mathcal{Q}$ method and guessing from the best dictionary resulted in successes very close to the optimal line. After 100 guesses these methods had compromised 795 users, in comparison to the 870 users compromised by guessing the correct password in the correct order for every guess.

\subsection{Password set 2: 60\% 000webhost, 30\% hotmail, 10\% computerbits}
\begin{figure}
    \centering
    \includegraphics[width=1\columnwidth]{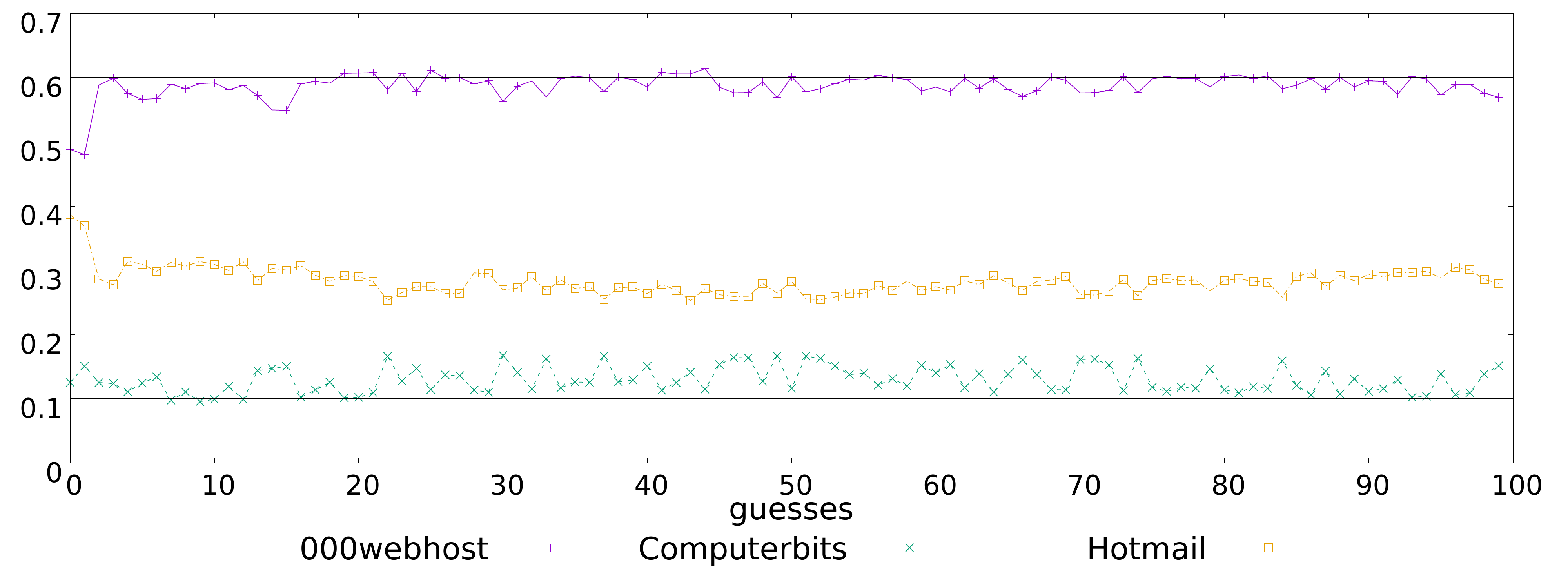}
        \vspace{-0.3cm}
    \caption{Password set 2 $q$-value estimates. Initialization: random $\hat q$-values, Guessing: random dictionary.}
    \label{fig:q-pwdset2}
\end{figure}

In Figure~\ref{fig:q-pwdset2}, we show the estimated $q$-values for a 10,000 user password set made from 000webhost, hotmail and computerbits with a 6:3:1 split. Again, we get very good estimates for the $q$-values. As the 000webhost passwords had composition rules in force but the other dictionaries did not, we may see different behaviour for the guessing successes.

\begin{figure}
    \centering
    \includegraphics[width=1\columnwidth]{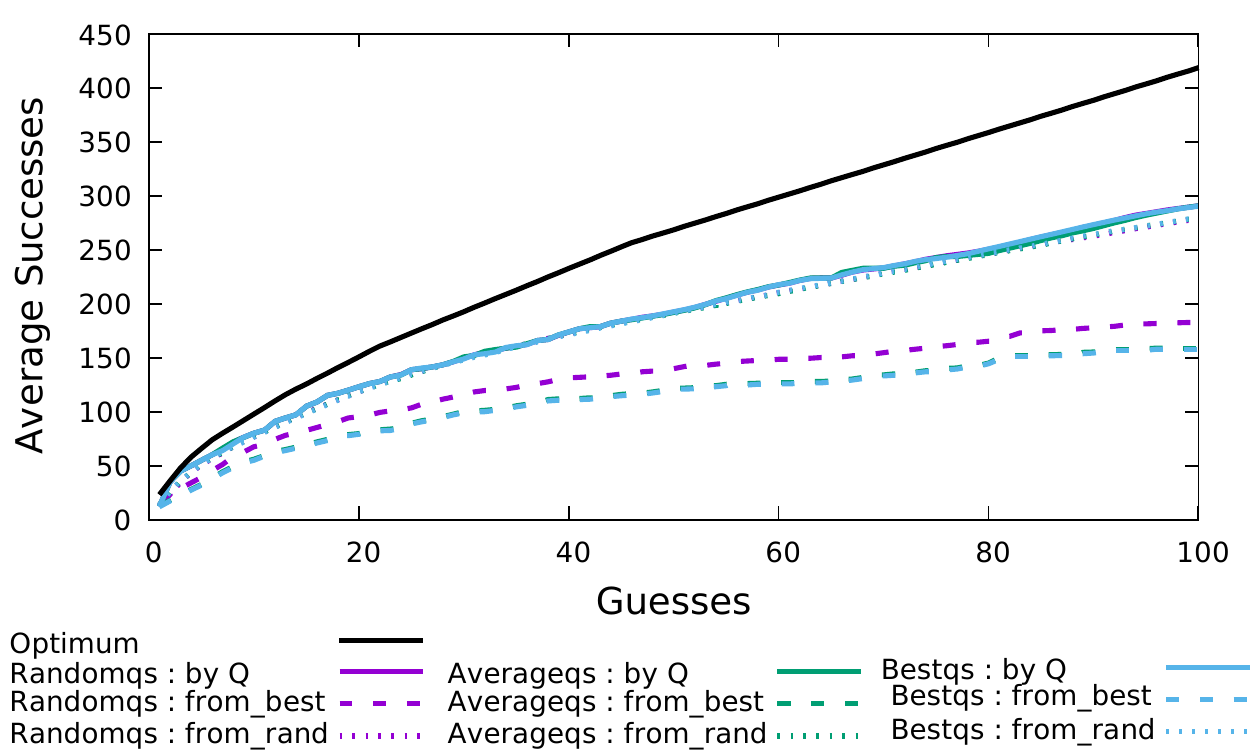}
    \caption{Guessing returns for password set 2.}
    \label{fig:success-pwdset2}
\end{figure}

In Figure~\ref{fig:success-pwdset2}, we show the guessing success rate. Interestingly, in this case, guessing from the best dictionary performed even worse than guessing from a random dictionary. This could be a reflection of the make-up of the 000webhost dataset. It was the only dataset which included rules on how passwords should be formatted~\cite{golla2018accuracy}. In our previous research~\cite{murray2020convergence} we found that this made it less effective at guessing. Here we are likely seeing a reflection of that. 

The $\mathcal{Q}$ method of guessing would also be skewed by the high ranking of the 000webhost passwords and their low guessing success. However, it still performs slightly better than the random method, and significantly better than guessing from the best dictionary (avg. results over 100 trials). 


Initialising with random $q$-values performs better than other initialisation methods when guessing using the best dictionary method. When $q$-values are initialised randomly, one dictionary can be ranked very high in comparison to the others and the gradient descent may not have been given sufficient time to converge. In this case, a password set other than 000webhost can be ranked as best. It is this that allows the random method to perform better than the other initialisation methods. 



\subsection{Password set 3: 55\% hotmail, 30\% flirtlife, 10\% 000webhost, 5\% computerbits}

\begin{figure}
    \centering
    \includegraphics[width=1\columnwidth]{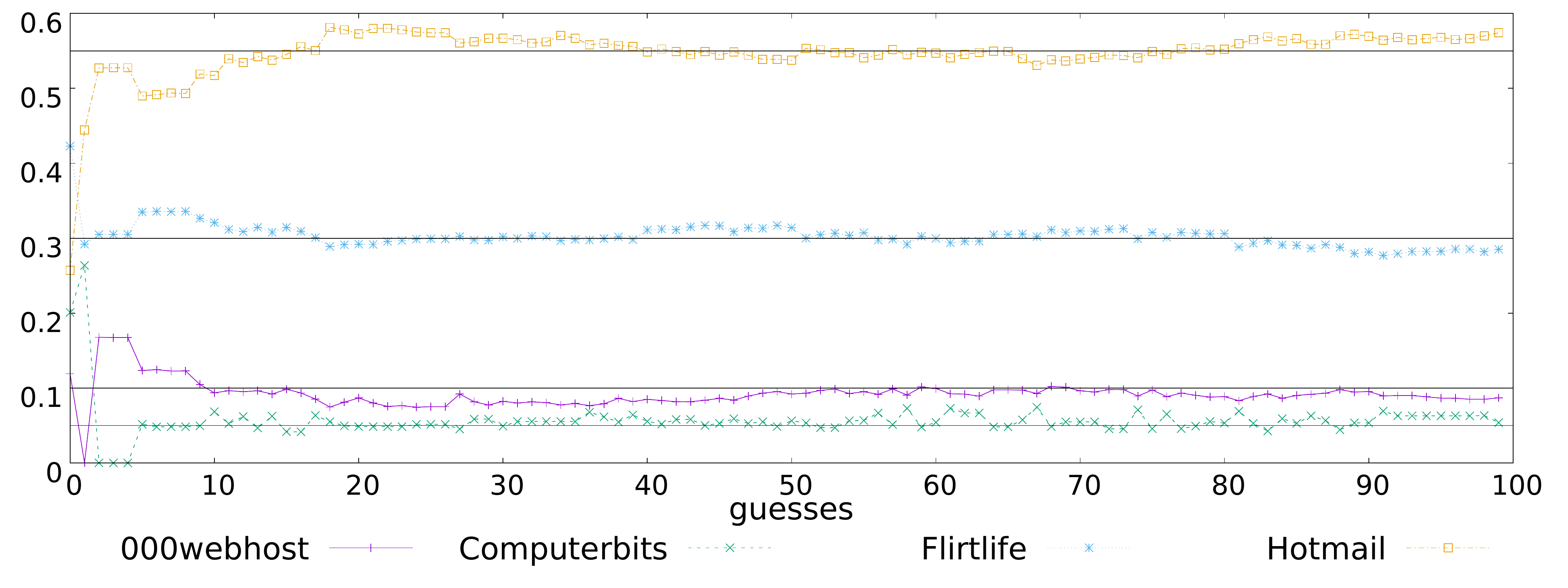}
        \vspace{-0.3cm}
    \caption{Password set 3 $q$-value estimates. Initialization: previous best $\hat q$-values, Guessing: random dictionary.}
    \label{fig:q-pwdset4}
\end{figure}

\begin{figure}
    \centering
    \includegraphics[width=1\columnwidth]{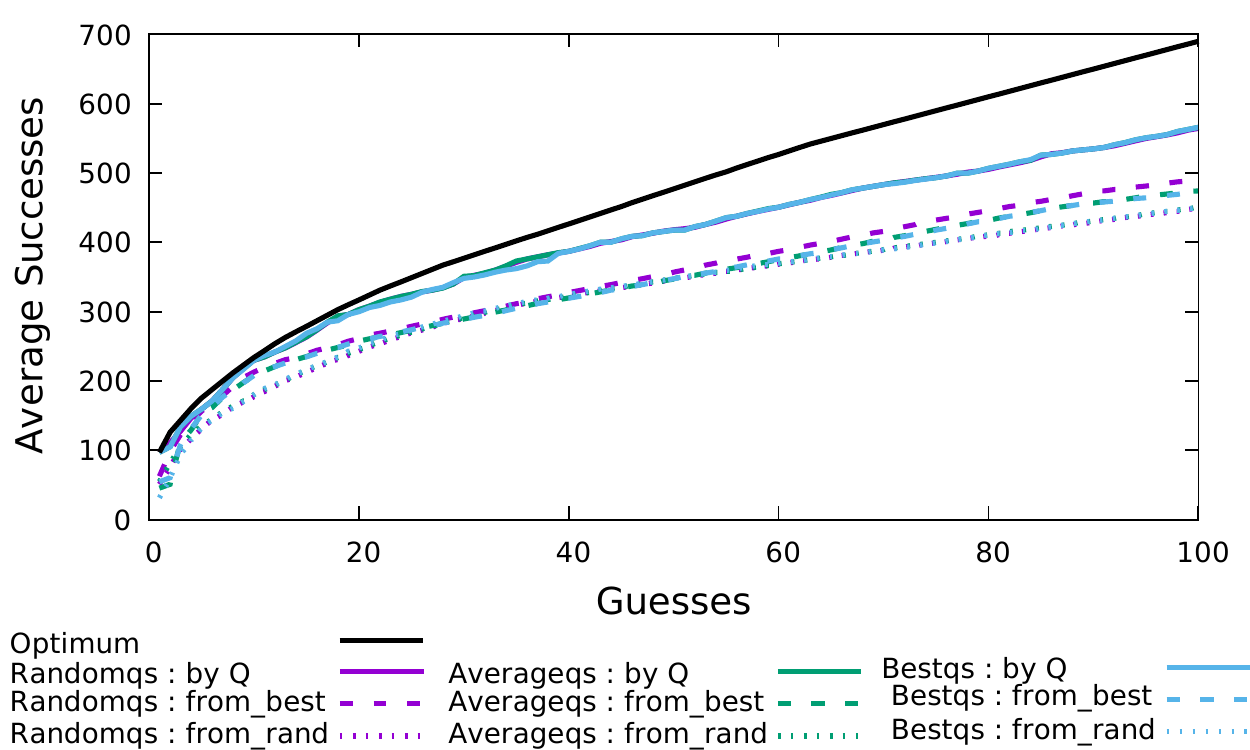}
    \vspace{-0.6em}
    \caption{Guessing returns for password set 3.}
    \label{fig:success-pwdset4}
\end{figure}

The final password set we look at is composed of 10,000 users' passwords from 4 different dictionaries. In Figure~\ref{fig:q-pwdset4}, we display the estimated $q$-values. Figure~\ref{fig:success-pwdset4} shows the successes when guessing this password set. Again, we see that the $\mathcal{Q}$ method is effective at guessing, this time performing significantly better than the other guessing methods. We notice that the successes are close to the optimal. Particularly for the first 20 guesses, where the $\mathcal{Q}$ method compromised 303 users in comparison to 317 compromised by optimal guessing. 

    \vspace{-0.2em}
    
\subsection{Discussion of Results}
The multi bandit automation is able to match characteristics in a passwordset to characteristics in the dictionaries used for guessing. We have seen that for a variety of synthetic examples, guessing using the multi-armed bandit technique can be effective both for compromising users and estimating how the passwords have been chosen.

In all examples we saw that guessing using the $\mathcal{Q}$ function is consistently effective in comparison to other dictionary selection methods. In general, we found that the initialization method had little bearing on the success results. This stems from the concave nature of our log-likelihood function, meaning that, for most set-ups, the function will converge to a single maximum when estimating the distributions. 

These initial results demonstrate that the relationship between password choice and user cohorts is tangible and identifiable by automation. This is potentially useful for both users and organizations. It provides further evidence for the importance of guiding users away from passwords which reflect characteristics associated with demographic or website specific terms. It also demonstrates that password choices differ measurably depending on their source use. This could indicate that websites could consider tailored blocklisting techniques. In particular, websites who have experienced previous password leaks could work at restricting future users from using passwords which occurred with a high frequency in that leak.





\section{Conclusion}
\label{sec:conclusion}

We used a multi-armed bandit approach to uncover the distribution of a password set and to optimise the order we chose passwords for guessing from dictionaries, thus improving the success of our guessing. In future work we plan to investigate these results in relation to guessing real leaked password datasets.


\section*{Acknowledgments}

 This publication is supported in part by a grant from Science Foundation Ireland (SFI), co-funded under the European Regional Development Fund under Grant 13/RC/2077. H.~Murray was supported by an IRC 2017 GOI Scholarship.

\bibliographystyle{plain}
\bibliography{lib.bib}

\begin{thebibliography}{10}

\bibitem{hashcat}
Hashcat.
\newblock \url{https://hashcat.net}.
\newblock Accessed: 9th june 2020.

\bibitem{jtr}
John the ripper password cracking.
\newblock \url{https://www.openwall.com/john/}.
\newblock Accessed: 9th June 2020.

\bibitem{castelluccia2013privacy}
C~Castelluccia, A~Chaabane, M~D{\"u}rmuth, and D~Perito.
\newblock When privacy meets security: Leveraging personal information for
  password cracking.
\newblock {\em arXiv:1304.6584}, 2013.

\bibitem{golla2018accuracy}
Maximilian Golla and Markus D{\"u}rmuth.
\newblock On the accuracy of password strength meters.
\newblock In {\em CCS '18}, pages 1567--1582, 2018.

\bibitem{houshmand2017using}
S~Houshmand and S~Aggarwal.
\newblock Using personal information in targeted grammar-based probabilistic
  password attacks.
\newblock In {\em IFIP Int. Conf. on Digital Forensics}, pages 285--303.
  Springer, 2017.

\bibitem{malone2012investigating}
David Malone and Kevin Maher.
\newblock Investigating the distribution of password choices.
\newblock In {\em Proceedings of the 21st international conference on World
  Wide Web}, pages 301--310, 2012.

\bibitem{morris1979password}
R~Morris and K~Thompson.
\newblock Password security: A case history.
\newblock {\em Comms. of the ACM}, 22(11):594--597, 1979.

\bibitem{murray2020convergence}
H~Murray and D~Malone.
\newblock Convergence of password guessing to optimal success rates.
\newblock {\em Entropy}, 22(4):378, 2020.

\bibitem{narayanan2005fast}
A~Narayanan and V~Shmatikov.
\newblock Fast dictionary attacks on passwords using time-space tradeoff.
\newblock In {\em CCS '05}, pages 364--372, 2005.

\bibitem{wei2018password}
M~Wei, M~Golla, and B~Ur.
\newblock The password doesn’t fall far: How service influences password
  choice.
\newblock {\em Who Are You}, 2018.

\bibitem{weir2009password}
M~Weir, S~Aggarwal, B~De~Medeiros, and B~Glodek.
\newblock Password cracking using probabilistic context-free grammars.
\newblock In {\em 30th IEEE Symp. on Security and Privacy}, pages 391--405.
  IEEE, 2009.

\end{thebibliography}

\end{document}